\documentclass[pra,twocolumn,superscriptaddress,10pt,noshowpacs]{revtex4}
\usepackage[english]{babel}
\usepackage[T1]{fontenc}
\usepackage[utf8]{inputenc}
\usepackage{graphicx,epstopdf}
\usepackage{amsmath}

\usepackage{amsfonts}
\usepackage{bbm}
\usepackage{amssymb}
\usepackage{color}
\usepackage{latexsym}
\usepackage{caption}
\usepackage{subcaption}
\usepackage{times,txfonts}

\begin{document}

\title{On the traversability of wormhole solutions in asymptotically safe gravity}

\author{M. Nilton}
\affiliation{Universidade Federal do Cear\'{a}, Fortaleza, Cear\'{a}, Brazil}
\email{matheus.nilton@fisica.ufc.br}

\author{J. Furtado}
\affiliation{Universidade Federal do Cariri(UFCA), Av. Tenente Raimundo Rocha, \\ Cidade Universit\'{a}ria, Juazeiro do Norte, Cear\'{a}, CEP 63048-080, Brasil}
\email{job.furtado@ufca.edu.br}

\author{G. Alencar}
\affiliation{Universidade Federal do Cear\'{a}, Fortaleza, Cear\'{a}, Brazil}
\email{geova@fisica.ufc.br}

\date{\today}

\begin{abstract}
In this paper we study the traversability of a wormhole in the context of Asymptotically Safe Gravity (ASG) for two different cases, namely, the spherical case and the pseudospherical case. We carried out an analysis in the throat of the wormhole with an specific choice of the cutoff function $f=\xi R$. Recently, this study has been performed in Ref \cite{Moti:2020whf}, however the authors consider only a classical source with constant state parameter $\omega$. Here we have generalized their study in two ways: a) with a classical state parameter which is dependent of the position $\omega(r)$ and b) with an ASG improvement of the source. We have showed that it is possible to have traversability with exotic and non-exotic matter, but the kind of matter is highly dependent on the wormhole's parameters.  
\end{abstract}

\maketitle

\section{Introduction}

\hspace{0.4cm}The idea of wormhole, i.e., a bridge connecting two asymptotically flat regions of the same universe or two different universes was first hypothesized in \cite{Einstein:1935tc} and it was known as Einstein-Rosen bridge. However, wormholes traversability was only studied more than fifty years later by Morris and Thorne \cite{Morris:1988cz}. An important feature of wormholes in Einstein's theory of gravity is that it's traversability requires that, in the case of an anisotropic fluid as source, we must have a radial pressure $p_r$ such that the dimensionless exoticity function $\xi_e=-(p_r+\rho)/\rho$ it must be positive, with $\rho$ being the energy density of the fluid \cite{Morris:1988cz}. This type of matter was called exotic and it is shown that leads a violation of the energy conditions \cite{Alcubierre:2017pqm}. This exotic matter content can be seen in a large number of wormholes being possible to have wormholes with phantom as energy source \cite{Sushkov:2005kj, Lobo:2005us} or even Casimir energy \cite{Garattini:2019ivd, Jusufi:2020rpw, Alencar:2021ejd, Oliveira:2021ypz, Carvalho:2021ajy}. Therefore the search for traversable wormholes in modified theories of gravity \cite{Richarte:2007zz, Matulich:2011ct, Richarte:2009zz, MontelongoGarcia:2011ag, Ovgun:2018xys, Chew:2016epf, Chew:2018vjp} without the requirement of exotic matter became an intense topic of research in the literature.

Among the modified theories of gravity, the Asymptotically Safe Gravity (ASG) has attracted a lot of attention recently due to fact that it presents a particularly interesting method for treating quantum effects \cite{Weinberg}. Furthermore, the ASG formalism proved to be a strong candidate for a quantum theory of gravity for having cured the problem of the curvature singularity of black-holes \cite{Bonanno:2000ep,Saueressig:2015xua,Pawlowski:2018swz,Platania:2019kyx,Bosma:2019aiu,Ishibashi:2021kmf}. It is also related to other methods for describing quantum gravity, such as the destructive interference of singular spacetimes in the Lorentzian path integral approach to quantum gravity and the Finite Action Principle \cite{Borissova:2020knn,Chojnacki:2021xtr,Giacchini:2021pmr,Chojnacki:2021ves}, where the ASG method motivates the presence of terms of curvature of higher-order in the action. In particular, we will apply the ASG method in the context of wormholes in order to verify if there is a possibility that they are traversable with non-exotic matter, contrary to what is predicted by classical theory.

The main idea in the ASG approach is to consider the quantum effects in gravity as asymptotically safe quantum field theory, which is UV complete. For this, we have to solve the exact renormalization group equation (ERGE) in order to determine the gravitational flow $\Gamma_k$. However, obtain exact solutions of the ERGE is a difficult task, and so, we have to appeal to the truncation method, that is, project the flow in a sub-space theory that accounts all the necessary physics, and then we extract the $\beta$-functions of the theory. The $\beta$-functions are a set of differential equations for all the running coupling constants that are essentials to describe the interactions in the theory. Once we have the 
$\beta$-functions, we must search for the fixed points. The theory is said asymptotically safe and free of UV divergences if exists a non-Gaussian fixed point such that the running coupling constants tend to it in the UV limit. There are several works verifying the existence of such a fixed point for the gravity renormalization group flow in various scenarios \cite{Reuter:1996cp, Lauscher:2001ya, Litim:2003vp, Machado:2007ea, Benedetti:2009rx, Manrique:2011jc, Christiansen:2012rx, Morris:2015oca, Demmel:2015oqa, Platania:2017djo, Christiansen:2017bsy, Falls:2018ylp, Narain:2009fy, Oda:2015sma, Eichhorn:2017ylw, Eichhorn:2019yzm, Reichert:2019car, Daas:2020dyo}. The minimal truncation is the Einstein-Hilbert truncation, which consists in associate the flow $\Gamma_k$ with the classical action for the gravitational field, with the gravitational coupling constants, i.e, the Newton's constant and the cosmological constant, becoming a function of the renormalization group scaling parameter $k$. Disregarding the cosmological influence this leads to the following form of the running coupling constant
\begin{equation}
G(k)=\frac{G_{0}}{1+\omega G_0 k^{2}},
\end{equation}
where $\omega=\frac{4}{\pi}(1-\pi^{2}/144)$ and $G_0$ is the measured value of the Newton's constant \cite{Reuter:1996cp,Bonanno:2000ep}. Therefore, the effects of gravity quantization in the ASG scenario can be effectively treated by turning the classical coupling constant into a running one, which emerges from the solution for the $\beta$-function \cite{Reuter:1996cp, Reuter:2003ca, Reuter:2004nx, Eichhorn:2021qet, Eichhorn:2022jqj}.

As was argued in \cite{Moti:2018rho}, for  gravitational systems the parameter $k$ must be identified as a function of 20 curvature invariants, denoted by $\chi$, corresponding to the 20 independent components of the Riemann tensor, which describes the tidal forces. This identification makes the geodesic deviation determine the scale of distances in space-time in an absolute way, so that small distances are defined as those in which the geodesic deviation is negligible. This considerations leads to an antiscreening running coupling constant \cite{Moti:2018rho}
\begin{equation}\label{Gchi}
    G(\chi)=\frac{G_0}{1+f(\chi)},
\end{equation}
where $\chi$ parameter is a function of the curvature invariants (such as $R$, $R_{\alpha\beta}R^{\alpha\beta}$, $R_{\alpha\beta\kappa\lambda}R^{\alpha\beta\kappa\lambda}$,...) and $f(\chi)=\xi/\chi$ is called anti-screening function, where $\xi$ is a dimensionless constant which measures the corrections due the ASG, in way that in the limit $\xi\rightarrow 0$ we recover the results predicted by General Relativity. 

With this, we can make the improvement by putting the running constant \eqref{Gchi} directly in the classical solutions or in the field equations, for example. However, as is argued in \cite{Moti:2018rho}, the more physical way is to introduce it directly into the action. Therefore, as a consequence of the quantum corrections emergent from ASG, the Einstein-Hilbert action is modified by the simply replacement of the Newton's constant by the improved coupling constant (\ref{Gchi}), providing the following form for the effective action in ASG context
\begin{equation}
S=\frac{1}{16\pi}\int\,d^{4}x\frac{\sqrt{-g}}{G(\chi)}R+S_{M},
\end{equation}

where $S_{M}$ is the action of the matter fields. With the action improvement the general covariance of the theory is saved and, in addition, the dynamics of the quantum corrections encoded in \eqref{Gchi} will appear naturally in the field equations. It consequently leads to quantum improved field equations of the form
\begin{equation}\label{ASG2}
    G_{\mu\nu}=8\pi G(\chi) T_{\mu\nu}+G(\chi)X_{\mu\nu},
\end{equation}
where $T_{\mu\nu}$ is the improved energy momentum tensor and $X_{\mu\nu}$ is a covariant tensor related to the derivation of $G(\chi)$ with respect to the metric \cite{Moti:2020whf}, which is given by
\begin{eqnarray}
\nonumber X_{\mu\nu}&=&\left(\nabla_{\mu}\nabla_{\nu}-g_{\mu\nu}\Box\right)G(\chi)^{-1}\\
\nonumber&&-\frac{1}{2}\left[R\mathcal{K}(\chi)\frac{\delta\chi}{\delta g^{\mu\nu}}+\partial_k\left(R\mathcal{K}(\chi)\frac{\partial\chi}{\partial(\partial_k g^{\mu\nu})}\right)\right.\\
&&\left.+\partial_k\partial_{\lambda}\left(R\mathcal{K}(\chi)\frac{\partial\chi}{\partial(\partial_{\lambda}\partial_k g^{\mu\nu})}\right)\right],
\end{eqnarray}
with $\mathcal{K}(\chi)=\frac{2}{G(\chi)^2}\frac{\partial G(\chi)}{\partial\chi}$. 

The tensor $X_{\mu\nu}$ describes the dynamics coming from the  the RG improved coupling
$G(\chi)$ and can be interpreted as an effective energy-momentum tensor $T^{eff}_{\mu\nu}$, related to the 4-momentum of the field $G(\chi)$. The above equation can be written  as
\begin{equation}
    \frac{1}{G(\chi)}\left(G_{\mu\nu}-T^{eff}_{\mu\nu}\right)=8\pi T_{\mu\nu}
\end{equation}
We should bear in mind that $T_{\mu\nu}$ is not classic. In order to properly consider the effects of ASG, we will also consider corrections to the matter sector\cite{Ishibashi:2021kmf,Moti:2021vck,Daum:2009dn,Christiansen:2017gtg}.  It is therefore carrying a parameter $\xi$ which comes from the improvement of the classical matter. Here we are not interested in an specific improvement of the matter sector and will consider a general case. As we will see below, this we enlarge the parameters of our system to four.

For non-vacuum solutions it is common to adopt a simple choice for $\chi$, such as $\chi=R^{-1}$ or $\chi=(R_{\alpha\beta}R^{\alpha\beta})^{1/2}$ \cite{Moti:2018rho, Moti:2019mws, Babic:2004ev, Domazet:2010bk, Domazet:2012tw}. The simple choice $\chi=R^{-1}$ is particularly important since we want to compare with the results obtained by Ref. \cite{Moti:2020whf}. For example, we will find other possibilities of non-exotic matter, not found in Ref. \cite{Moti:2020whf}, when $\chi=R^{-1}$. Of course, if we consider other possibilities of $\chi$ new results should be achieved. However we let this for future studies. We also point that this model can also be compared to other modified gravity theories, such as $f(R)$ \cite{Hindmarsh:2012rc}.

In the context of ASG, several studies were carried out in the attempt of investigating the effects of quantum improvement in compact objects, such as black-holes \cite{Ruiz:2021qfp, Liu:2012ee, Rincon:2020iwy} and in the traversability of wormholes. In \cite{Moti:2020whf}, the authors have shown that improved pseudospherical wormholes could be traversible with nonexotic matter, while spherical ones could not. In \cite{Alencar:2021enh}, the authors investigated the Ellis-Bronnikov wormhole solution in the ASG context and it was showed that, for a perfect fluid, it is not possible to get a traversible Ellis-Bronnikov wormhole without exotic matter, and in order to be traversible the wormhole requires very exotic phantom-like matter in some regions of the modified spacetime and even at the throat. And more recently, it was showed in \cite{Nilton:2021pyi} that Schwarzschild-like wormholes exhibit the possibility of having nonexotic matter as source for certain values of the radial coordinate.

In this paper we study the traversability of a wormhole in the context of asymptotically safe gravity for two different cases, namely, the spherical case and the pseudospherical case. We carried out an analysis in the throat of the wormhole with an specific choice of the cutoff function $f=\xi R$. We have also considered a position dependent equation of state in a zero tidal model aiming to investigate if it is possible to have a traversable wormhole with non-exotic matter in the context of ASG.  

This paper is organized as follows: In the next section we discuss the traversability of a pseudospherical wormhole in the context of ASG for a non-linear equation of state in the throat of the wormhole. In section III we perform the same analysis for a spherical wormhole and in section IV we present our conclusions.

\section{The Pseudospherical case}
We begin this discussion with the Pseudospherical wormholes, a class of hyperbolic wormholes whose metric is given by the ansatz
\begin{equation}
    ds^2 = e^{2\Phi(r)} dt^2 - \frac{dr^2}{1-b(r)/r} -r^2d\Omega_{2(p)}, \label{PSM}
\end{equation}
where $e^{2\Phi}$ is the redshift function, $b(r)$ the shape function and $d\Omega_{2(p)}=d\theta^{2}+\sinh^{2}{\theta}d\phi^{2}$ is the line element of a 2-pseudosphere, being important in the context of the spacetimes with the negative curvature, being a kind of ``anti-Morris-Thorne metric'' analogous to de Sitter and anti-de Sitter spacetimes. In particular, as it shown in \cite{Cataldo:2015vra}, this class of wormholes in the context of general relativity leads a negative energy density and a positive radial pressure at the throat, contrary to what happens in the Morris-Thorne usual case.

Considering an anisotropic fluid $T^{\mu}_{\nu} = \text{Diag} [\rho,-p_r,-p_l,-p_l]$ as source to the wormhole spacetime, the improved field equations for the metric \eqref{PSM} provides the following modified field equations \cite{Moti:2020whf}
\begin{eqnarray}
    \nonumber 8\pi G_0 \rho &=&\bigl(1+f\bigr) \frac{b^{'}-2}{r^2} -\left(1-\frac{b}{r}\right) \left(f''+\frac{2}{r} f'\right)\\
    &&+ \frac{b^{'}r-b}{2r^2}f'  \label{IEQ-tt-p}\\
    \nonumber 8\pi G_0 p_r  &=& -\bigl(1+f\bigr) \left(\frac{b}{r^3} -\frac{2}{r^2} -\frac{2\Phi^{'}}{r}\left(1-\frac{b}{r}\right) \right) \\
    &&+ \left(1-\frac{b}{r}\right) \left( \Phi^{'}+\frac{2}{r}\right) f'   \label{IEQ-rr-p} \\
    \nonumber8\pi G_0 p_l  &=& -\bigl(1+f\bigr) \left( \frac{b'r-b}{2r^2} \left(\Phi'+\frac{1}{r} \right)\right. \\
    \nonumber&&-\left. \left(1-\frac{b}{r}\right)\left(\Phi''+\Phi'^2+\frac{\Phi'}{r}\right) \right)\\
     &&+ \left(\left(1-\frac{b}{r}\right) \left( \Phi^{'}+\frac{1}{r}\right)- \frac{b^{'}r-b}{2r^2}\right) f' + \left(1-\frac{b}{r}\right)f'' \ , \label{IEQ-pp-p}
  \end{eqnarray}
where $\rho$ is the energy density, $p_{r}$ and $p_{l}$ are the radial and lateral pressures, respectively, and the prime stands for the derivative with respect to $r$.

We will consider, initially, the simplest choice for the anti-screening $f$ function in terms of the Ricci scalar, thus we define $f=\xi R$. Furthermore, we will consider that the anisotropic matter threading the wormhole obeys a state equation of the form $p_{r}(r,\xi)=\omega(r,\xi)\rho(r,\xi)$, where $\rho$ and $p_r$ are respectively given by the Eqs. \eqref{IEQ-tt-p} and \eqref{IEQ-rr-p} and $\omega(r,\xi)$ is the position-dependent state parameter that determines whether the matter is exotic or not as a position-function. Note that the parameter $\xi$ is a scaling constant, so that the usual General Relativity is recovered when $\xi\rightarrow 0$. Using the equations \eqref{IEQ-tt-p} and \eqref{IEQ-rr-p} the equation of state we have
\begin{eqnarray}
\label{linear-eq-p}
     \nonumber&&\omega b' r +b -2(\omega+1)r  -2 \Phi' r^2 \left(1-\frac{b}{r}\right)=\\
     \nonumber&&\frac{r^3}{1+f} \left(\left(1-\frac{b}{r}\right)(\omega f''+\frac{2}{r} (1+\omega)f'+\Phi'f')-\omega\frac{b^{'}r-b}{2r^2}f'  \right).
\end{eqnarray}

Furthermore, we consider a wormhole without tidal-forces putting $\Phi(r)=0$. The zero-tidal force consideration is interesting since a stationary observer travelling through the wormhole should not see a tidal force greater than Earth's gravity \cite{Morris:1988cz}. Also, the constant travel speed consideration through the wormhole leads naturally to the zero tidal force conditions. Besides, we define the dimensionless radial coordinate $u \equiv r/r_{t}$, where $r_{t}$ is the throat radius, so that under these conditions the equation $\eqref{linear-eq-p}$ becomes
\begin{eqnarray}
  \nonumber\omega \dot{\tilde{b}} u +\tilde{b} -2(\omega+1)u &=& \frac{u^3}{1+f} \left(\left(1-\frac{\tilde{b}}{u}\right)\left(\omega \ddot{f}+\frac{2}{u} (1+\omega)\dot{f}\right)\right.\\
  &&\left.-\omega\frac{\dot{\tilde{b}}u-\tilde{b}}{2u^2}\dot{f}  \right), \label{be-p}
\end{eqnarray}
where $\tilde{b} \equiv b/r_{t}$ and the dot represents the derivative with respect to $u$. 

Our objective is to analyze the influence of ASG to the energy conditions and to the state parameter, given by $\omega=p_r(r,\xi)/\rho(r,\xi)$. As said before, $\rho(r,\xi)$ and $p(r,\xi)$ depends on $\xi$ due to the improvement of ASG.  In the classical limit $\xi\rightarrow 0$, $f = 0$ and Eq. \eqref{be-p} recover the general relativity case. In this sense, the classical solution is obtained only in the limit $\xi\to 0$ and we get $b_{(0)}$ and $p_{(0)}=\omega_{(0)}\rho_{(0)}$ where $\rho_{(0)}=\rho(r,0)$ and so on. In order to get the influence of ASG we departure from this zeroth order solution and solve the above equation interactively as proposed in Ref.\cite{Moti:2020whf}. We consider an expansion of the form $\tilde{b}(u) \approx \tilde{b}_{(0)}(u)+\tilde{b}_{(1)}(u)$, where the term of zeroth order is the solution of the equation \eqref{be-p} when we have no ASG contribution ($\xi=0$). Also, in order to consider the most general case, i.e., an arbitrary state parameter $\omega(u,\xi)$, we must perform a similar expansion for the state parameter as $\omega(u)=\omega_{(0)}(u)+\omega_{(1)}(u,\xi)$, where the zeroth order is the contribution independent of ASG.

Such expansion for the state parameter is reasonable since from Einstein's equations (\ref{IEQ-tt-p}) and (\ref{IEQ-rr-p}) we can see that the energy density and radial pressure can also be separated in an ASG independent part and dependent one and write
\begin{eqnarray}
    \rho&=&\rho_{(0)}(u)+\rho_{(1)}(u,\xi)\\
    p_r&=&p_{r(0)}(u)+p_{r(1)}(u,\xi),
\end{eqnarray}
where $\rho_{(0)}$ and $p_{r(0)}$ are the usual General Relativity contributions for the energy density and radial pressure respectively and $\rho_{(1)}$ and $p_{r(1)}$ are the contributions due the ASG. This highlights the fact that the corrections due to ASG can be treated by considering an effective energy momentum tensor. Therefore, the state parameter $\omega$
\begin{eqnarray}
    \nonumber\omega(u)&=&\frac{p_r}{\rho}=\left[\frac{p_{r(0)}+p_{r(1)}}{\rho_{(0)}}\right]\left(1+\frac{\rho_{(1)}}{\rho_{(0)}}\right)^{-1}\approx\left[\frac{p_{r(0)}}{\rho_{(0)}}+\frac{p_{r(1)}}{\rho_{(0)}}\right]\left(1-\frac{\rho_{(1)}}{\rho_{(0)}}\right)\\
    &=&\omega_{(0)}(u)+\omega_{(1)}(u,\xi),
\end{eqnarray}
also presents naturally a decomposition into GR and ASG parts. It is important to note that in the classical limit $\xi\rightarrow 0$ we will also have a linear equation of state $p_{r(0)}=\omega_{(0)}(u)\rho_{(0)}$ with a state parameter $\omega_{(0)}(u)$ dependent on the position, being a more general case than that considered in Ref.\cite{Moti:2020whf}, where the authors consider only the special solution $\omega_{(0)}=constant$ of General Relativity. Therefore, when $\xi=0$ we have
\begin{equation}
\label{zero-order-eq}
\omega_{(0)}(u) \dot{\tilde{b}}_{(0)} u +\tilde{b}_{(0)} -2(\omega_{(0)}(u)+1)u = 0.
\end{equation}

However, since we are considering a general position-dependent state parameter $\omega_{(0)}=\omega_{(0)}(u)$, and we restrict ourselves to regions close to the throat, we will expand it around the $u=1$ so that
\begin{equation}
(\omega_t+\kappa(u-1)) \dot{\tilde{b}}_{(0)} u +\tilde{b}_{(0)} -2((\omega_t+\kappa(u-1))+1)u = 0,
\end{equation}
where we define the parameters $\kappa=d\omega_{(0)}(u)/du|_{u=1}$ and $\omega_t=\omega_{(0)}(1)$. Note that once we have a general equation of state in the GR limit, $p_{r(0)}=\omega_{(0)}(u)\rho_{(0)}$, the parameters $\kappa$ and $\omega_t$ arising from the linearization of $\omega_{(0)}$ are free. The solution is easily solved and gives
\begin{eqnarray}
\tilde{b}_{(0)}(u)=2 u-\omega_t^{\frac{1}{\kappa-\omega_t}} u^{\frac{1}{\kappa-\omega_t}} (\kappa (u-1)+\omega_t)^{\frac{1}{\omega_t-\kappa}}
\end{eqnarray}
for the initial condition $\tilde{b}_{(0)}(1) = 1$. Therefore, the zeroth order solution already depends on two parameters.

In order to obtain the differential equation for the $\tilde{b}_{(1)}$ term, we have to use the expansions $\tilde{b}\approx\tilde{b}_{(0)}+\tilde{b}_{(1)}$ and $\omega\approx\omega_{(0)}+\omega_{(1)}$ in Eq. \eqref{be-p}, which give us the following equation 
\begin{widetext}
\begin{eqnarray}
&\left[\omega_{(0)}\dot{\tilde{b}}_{(0)}u +\tilde{b}_{(0)} -2(\omega_{(0)}+1)u\right]+ \omega_{(0)}\dot{\tilde{b}}_{(1)}u+\tilde{b}_{(1)}+\omega_{(1)}\dot{\tilde{b}}_{(1)}u+\omega_{(1)}(\dot{\tilde{b}}_{(0)}-2)u &=\nonumber
\\ 
&\frac{u^3}{1+f} \left((1-\frac{\tilde{b}_{(0)}+\tilde{b}_{(1)}}{u})((\omega_{(0)}+\omega_{(1)}) \ddot{f}+ \frac{2}{u}(1+(\omega_{(0)}
+\omega_{(1)}))\dot{f})-(\omega_{(0)}+\omega_{(1)})\frac{(\dot{\tilde{b}}_{(0)}+\dot{\tilde{b}}_{(1)})u-(\tilde{b}_{(0)}+\tilde{b}_{(1)})}{2u^2}\dot{f}\right).& \label{complete-eq}
\end{eqnarray}
\end{widetext}
In the left-hand side of the Eq.\eqref{complete-eq}, the term in parenthesis is exactly the zero--order term, which naturally vanishes due the Eq. \eqref{zero-order-eq}, and we will disregard the product $\omega_{(1)}\tilde{b}_{(1)}$ since it is a higher-order term in $\xi$. Now, due the presence of the terms of derivative of $f$ in the right-hand side of the Eq. \eqref{complete-eq}, all the terms that involve order 1 will be negligible  because the linearity in $\xi$ of $f$ and therefore it gives terms of higher-order in $\xi$.

Thus, the term of the first order is obtained substituting the $\tilde{b}_{(0)}$ and $\omega_{(0)}$ on the right side of \eqref{be-p} and we get
\begin{widetext}
\begin{equation}
\label{eq-b1}
\omega_{(0)}\dot{\tilde{b}}_{(1)}u+\tilde{b}_{(1)}+\omega_{(1)}(\dot{\tilde{b}}_{(0)}-2)u=\frac{u^{3}}{1+f}\left(\left(1-\frac{\tilde{b}_{(0)}}{u}\right)\left(\omega_{(0)}\ddot{f}+\frac{2}{u}(1+\omega_{(0)})\dot{f}\right)-\omega_{(0)}\frac{\dot{\tilde{b}}_{(0)}u-\tilde{b}_{(0)}}{2u^{2}}\dot{f}\right).
\end{equation}
\end{widetext}
Note that the $\tilde{b}_{(1)}$, and therefore the complete shape function $\tilde{b}$, will not only depend on the $\omega_{(0)}$, but also depends of $\omega_{(1)}$. Therefore $\omega_{(0)}$ and $\omega_{(1)}$ can not be determined by the dynamics, but enter as unknown functions. However we can repeat the procedure to obtain $b_{(0)}$ to solve Eq. \eqref{eq-b1} near the throat. For this we expand $\omega_{(0)}$ and $\omega_{(1)}$ around the throat as $\omega_{(1)}=\omega_T\zeta+\lambda\zeta(u-1)$, where $\omega_T=\omega_{(1)}(1)$ and $\lambda=d\omega_{(1)}(u)/du|_{u=1}$. Therefore we get two more parameters $\omega_T$ and $\lambda$. With this, we can see that the shape function will have $4$ free parameters, two due the linearization of $\omega_{(0)}$ ($\omega_t$ and $\kappa$) and two due the linearization of $\omega_{(1)}$($\omega_T$ and $\lambda$). Thus, in what follows we will scan over the admissible values for the matter dynamics by analyzing the wormhole traversability in terms of the values of these 4 parameters.

With the above considerations, we use $\omega_{(0)}(u)=\omega_t+\kappa(u-1)$ in the anti-screening function using the Ricci Scalar. We get
\begin{equation}
f = \xi R = -2 \zeta  \omega_t^{\frac{1}{\kappa-\omega_t}} u^{\frac{1}{\kappa-\omega_t}-3} [\kappa (u-1)+\omega_t]^{\frac{1}{\omega_t-\kappa}-1},    
\end{equation}
where $\zeta=\xi/r_t^2$, and this provide for $\tilde{b}_{(1)}$ the following equation
\begin{widetext}
\begin{eqnarray}\label{b1psw}
\nonumber\omega_{(0)}\dot{\tilde{b}}_{(1)} u +\tilde{b}_{(1)}+\omega_{(1)}\left(\dot{\tilde{b}}_{(0)}-2\right)&=&\frac{\zeta \omega_t^{\frac{1}{\kappa-\omega_t}} u^{\frac{1}{\kappa-\omega_t}} (\kappa (u-1)+\omega_t)^{\frac{1}{\omega_t-\kappa}-1}}{u^3 (\kappa (u-1)+\omega_t)-2 \zeta  \Lambda}\left\{2 u \left(2 \kappa^2 \left(6 u^2-8 u+3\right)+4 \kappa (4 u-3) \omega_t+\kappa+6 \omega_t^2-\omega_t-1\right)\right.\\
&&\left.-\Lambda\left[\kappa^2 (u (28 u-39)+15)-1+2\omega_t+15\omega_t^2-30\omega_t-2+\kappa (u (39\omega_t+5))\right]\right\},
\end{eqnarray}
\end{widetext}
where for question of brevity we omit the expansion for $\omega_{(1)}$ in the left-hand side of the Eq. \eqref{eq-b1} and $\Lambda$ is defined as
\begin{eqnarray}
    \Lambda=\omega_t^{\frac{1}{\kappa-\omega_t}} u^{\frac{1}{\kappa-\omega_t}} (\kappa (u-1)+\omega_t)^{\frac{1}{\omega_t-\kappa}}.
\end{eqnarray}
The solution for $\tilde{b}_{(1)}$ and consequently for $\tilde{b}(u)$ are expressed in terms of hypergeometric $_2F_1$ functions and too lengthy to be reported here.

Hence we need to study how the wormhole's traversability is affected under different configuration of such parameters. Also, differently from the analysis performed in \cite{Alencar:2021enh, Nilton:2021pyi}, in this case it is not possible to write the expressions for pressure and energy density in terms of the wormhole parameters without specifying a value for the free parameters, since the shape function $\tilde{b}(u)$ presents an explicit dependence on them. 

As a first case, let us consider the configuration where $\omega_T=\kappa=\lambda=0$ and $\omega_t\neq0$. In this case the shape function $\tilde{b}(u)$ recovers the result presented in \cite{Moti:2020whf} and consequently the discussions on the wormhole's traversability follows what is reported in \cite{Moti:2020whf}. However such choice imposes severe restrictions on the system, since in this particular case we are working with a constant and ASG independent state parameter $\omega$.

Now, we will investigate the presence of regions with exotic matter. For this, we will apply the traversability conditions and see what restrictions these conditions bring to the parameters:
\begin{itemize}
    \item Throat Condition: To guarantee the existence of the throat, we need the following conditions: No event horizon, the condition of a minimum $b(r_{t})=r_{t}$, and the flare-out condition $b'(r_{t})<r_{t}$.
    
Since we are considering a zero-tidal model it is straightforward to see that there will be no event horizons. Furthermore, we already had the minimum condition, $\tilde{b}(1)=1$, by construction. Therefore, only the flare-out condition remains, $\dot{\tilde{b}}(1)<1$, which gives for our model
\begin{widetext}
\begin{equation}
\frac{-\zeta  [\kappa (\omega_t+1)+\omega_t (7 \omega_t+\omega_T+6)+1]+(2 \omega_t+1) \omega_t^2+2 \zeta ^2 \omega_T}{\omega_t^2 (\omega_t-2 \zeta )}-1<0.
\end{equation}
\end{widetext}
Note that the flare-out condition establishes a connection between $\omega_t$, $\omega_T$, $\zeta$ and $\kappa$ and it diverges when $\omega_t=2\zeta$. As we can see from fig.(\ref{fout1}), when $\omega_T\neq0$ we have a shift in the smallest root of the flare-out condition function, which apparently sets a small region where phantom-like matter is permitted by the flare-out condition. However we must remember that the state parameter on the throat is written as $\omega=\omega_t+\zeta\omega_T$, so that for the chosen parameters we have $\omega=-0.88$ (quintessense-like matter) for $\omega_t=-1.28$, which is the smallest root for the flare-out condition in fig.(\ref{fout1}).

\begin{figure}
    \centering
    \includegraphics[scale=0.9]{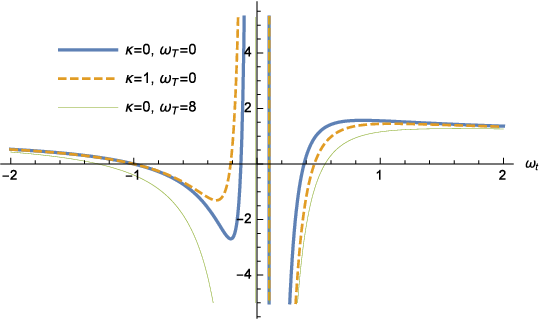}
    \caption{Flare-out condition for pseudospherical wormhole in ASG context with $\zeta=0.05$. Note that the flare-out condition is satisfied in the region where ordinary matter is permited and this leads a region with exotic quintessence-like matter.}
    \label{fout1}
\end{figure}

\item Antiscreening condition: this condition is entirely related to ASG. The exact renormalization group equation determines the dynamics of the running coupling constant $G(\chi)$, however the functional renormalization group methods along with other assumptions \cite{Reuter:1996cp, Souma:1999at} requires that we must have $f(\chi)|_{u \rightarrow u_{0}} > 0$, which gives for our model
\begin{equation}
\frac{2 \zeta  \left(\zeta  \omega_t (\kappa+\omega_T+6)+\zeta  (\kappa-2 \zeta  \omega_T+1)+(3 \zeta -1) \omega_t^2\right)}{\omega_t^2 (\omega_t-2 \zeta )}>0.
\end{equation}
Note that the antiscreening condition is important to guarantee the antiscreening behavior of gravity, that is, the running coupling constant decreasing with the energy. Similarly to the flare-out condition, the antiscreening condition also exhibits a divergence when $\omega_t=2\zeta$, but does not exclude the possibility of traversability with phantom-like matter.

\begin{figure}
    \centering
    \includegraphics[scale=0.9]{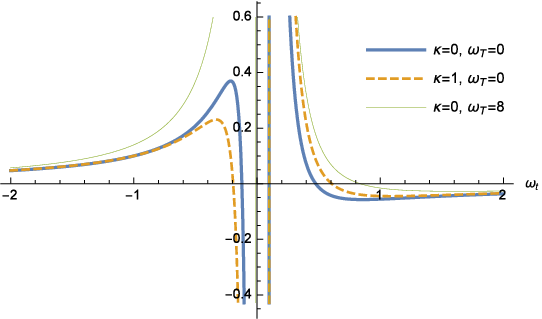}
    \caption{Antiscreening condition for pseudospherical wormhole in ASG context with $\zeta=0.05$. Note that in this case we also have an asymptote in $\omega_t=0.1$ as well as the flare-out condition and a large region with phantom matter is allowed in this condition. Again there is the possibility of the presence of ordinary matter in this condition.}
    \label{asc1}
\end{figure}

\begin{figure}[ht]
    \centering
    \includegraphics[scale=0.9]{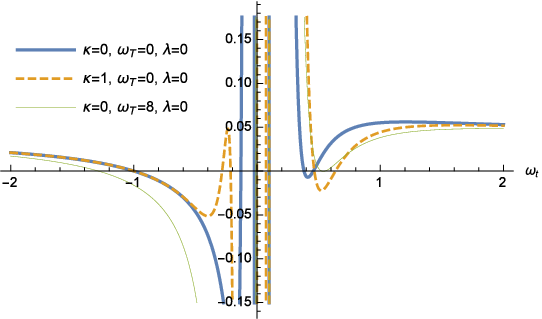}
    \caption{$p_r+\rho$ for pseudospherical wormhole in ASG context with $\zeta=0.05$. This condition also allow ordinary matter to have traversability and also the possibility of phantom-like matter. However the presence of phantom matter is have to be analyzed carefully as in done in flare-out condition.}
    \label{nec1}
\end{figure}

\begin{figure}[ht]
    \centering
    \includegraphics[scale=0.9]{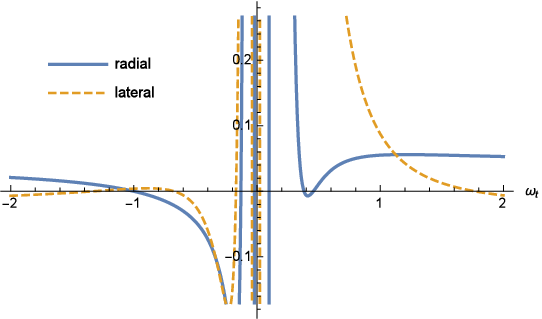}
    \caption{$p_l+\rho$ and $p_r+\rho$ for pseudospherical wormhole in ASG context with $\zeta=0.05$, $\kappa=0$, $\omega_T=0$ and $\lambda=0$. This plot presents a comparison between radial and lateral null energy conditions. We can see that, in the pseudospherical case, both radial and lateral null energy conditions have a similar behaviour being the radial one a bit more restrictive.}
    \label{necL1}
\end{figure}

\item The null energy condition: The null energy condition is important to guarantee the absence of Phantom matter as a source. For a zero--tidal model it can be written directly from Einstein equations (\ref{IEQ-tt-p}), (\ref{IEQ-rr-p}) and (\ref{IEQ-pp-p}), in which we substitute the shape function $\tilde{b}(u)$ and the cutoff function $f(u)$. Therefore, the radial null energy condition, i.e. $p_r+\rho>0$, can be written as presented in (\ref{neccond}). Similarly to the previous conditions, the radial null energy condition it is also divergent when $\omega_t=2\zeta$.

Note that differently from the throat and antiscreening condition, the radial null energy condition presents an explicit dependence on the $\lambda$ parameter. Also we can see from fig.(\ref{nec1}) that the asymptotic behaviour of the null energy condition is similar to the behaviour presented by the flare-out condition, so that the apparent permission of phantom-like matter must be analysed carefully, as we did previously.

Also, the lateral null energy condition, i.e. $p_l+\rho>0$ must be similarly analyzed. Analogously to the radial null energy condition, in this case we also have a divergence when $\omega_t=2\zeta$. The proper expression for the lateral null energy condition was omitted here for the sake of simplicity, however its behaviour is presented in fig. (\ref{necL1}). In general the radial and lateral null energy conditions present similar behaviour, at least in the pseudospherical case, being the radial null energy condition a bit more restrictive.

\begin{widetext}
\begin{eqnarray}\label{neccond}
    \nonumber\frac{1}{8 \pi  \omega_t^5 (\omega_t-2 \zeta )^3}&&\left[-\zeta  ((\omega_t+1) (\kappa+5 \omega_t+1)+\omega_t \omega_T)+(\omega_t+1) \omega_t^2+2 \zeta ^2 \omega_T\right]\times\\
    \nonumber&&\times\left\{-2 \zeta ^2 \omega_t^2 \left[2 \zeta  (\lambda -11)+3 \kappa^2+\kappa (-8 \zeta +\omega_T+14)-2 \zeta  \omega_T+\omega_T+8\right]\right.\\
    \nonumber&&+\zeta ^2 \omega_t \left((10 \zeta -7) \kappa^2+4 \kappa (\zeta  (2 \omega_T+9)-2)+2 \zeta  (2 \zeta  \lambda -2 (\zeta -2) \omega_T+7)-1\right)+\\
    &&\left.\zeta \omega_t^3 (-14 \zeta  \kappa+\kappa+\zeta  (30 \zeta +\lambda -\omega_T-35)+1)+4 \zeta ^3 (\kappa+1) (3 \kappa-2 \zeta  \omega_T)+\omega_t^5-3 \zeta  (8 \zeta +1) \omega_t^4\right\}>0
\end{eqnarray}
\end{widetext}

\end{itemize}

Finally, in order to properly study the traversability in the present context, we need to investigate the regions in the space of parameters where all the above conditions are satisfied simultaneously. 

\begin{figure*}[ht]
\centering
\begin{subfigure}{.3\textwidth}
  \centering
  \includegraphics[scale=0.52]{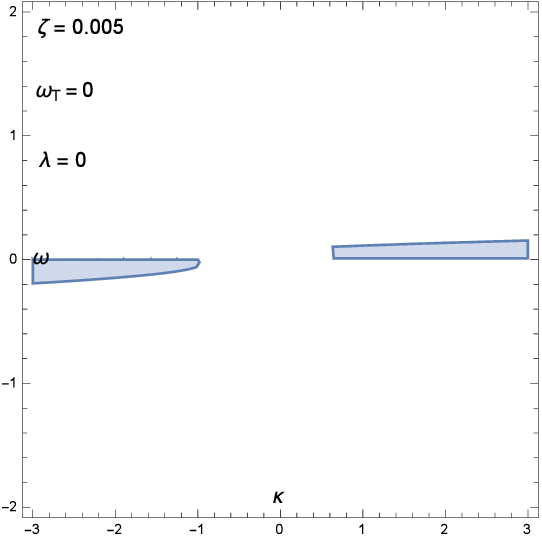}
  \caption{}
 
\end{subfigure}%
\begin{subfigure}{.3\textwidth}
  \centering
  \includegraphics[scale=0.52]{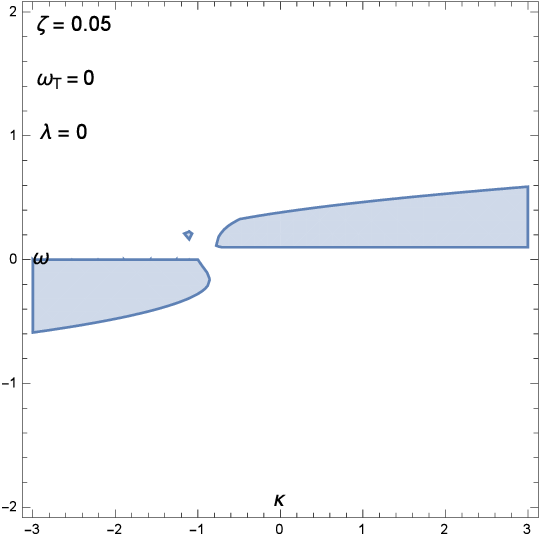}
  \caption{}

\end{subfigure}%
\begin{subfigure}{.3\textwidth}
  \centering
  \includegraphics[scale=0.52]{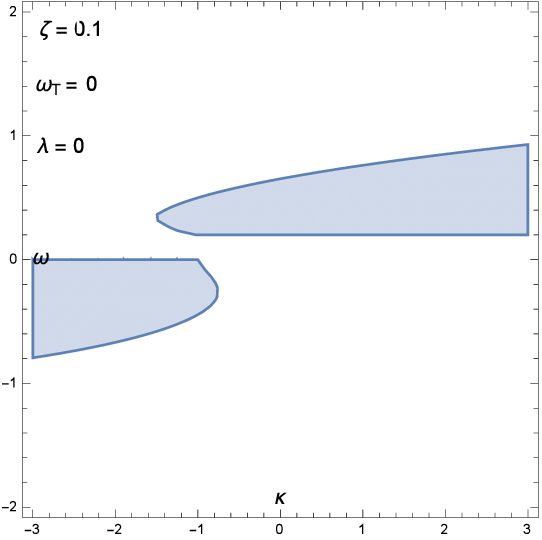}
  \caption{}
  
\end{subfigure}
\caption{In (a) we see the traversability region for pseudospherical worhmholes in the ASG context for $\zeta=0.005$, in (b) we have set $\zeta=0.05$ while in (c) $\zeta=0.1$. In all cases $\omega_T=\lambda=0$. Note that the traversability is allowed with nonexotic matter at throat as well as reported in \cite{Moti:2020whf}. However, we have the possibility to have traversability with quintessence-like matter.}
\label{travpsw}
\end{figure*}

In figure (\ref{travpsw}) we can see the region of validity in the space of parameters where all the conditions are simultaneously satisfied at the throat of the wormhole. We have set $\omega_T=\lambda=0$ in order to disregard the influence of ASG in the state parameter. In fig. (\ref{travpsw}a) we have considered $\zeta=0.005$ and we can see that around $\kappa=0$ there is no traversability. However by increasing the value of $\kappa$ traversability becomes allowed with ordinary matter, while by decreasing the value of $\kappa$ the traversability is only possible with exotic matter ($\omega_t<0$). We can also see, by comparing figures (\ref{travpsw}a), (\ref{travpsw}b) and (\ref{travpsw}c), that the increase in the value of $\zeta$ leads to an increase of traversable regions. This dependence on $\zeta$ is expected, since the greater is the $\zeta$ the greater is the quantum influence in the wormhole.  

\section{The Spherical Case}
Now, we turn our attention to the spherical case, whose metric is given by
\begin{equation}
   ds^2 = e^{2\Phi(r)} dt^2 - \frac{dr^2}{1-b(r)/r} -r^2d\Omega_{2(s)} \ , \label{SSM}
\end{equation}
where now $d\Omega_{2(s)} = d\theta^{2} + \sin^{2}{\theta}d\phi^{2}$ is the line element of a 2-sphere.

Again, considering an anisotropic fluid that generates the wormhole spacetime we obtain the modified field equations 
\begin{eqnarray}
    \nonumber 8\pi G_0 \rho &=&\bigl(1+f\bigr) \frac{b^{'}}{r^2} -\left(1-\frac{b}{r}\right) \left(f''+\frac{2}{r} f'\right)\\
    &&+ \frac{b^{'}r-b}{2r^2}f'  \label{IEQ-tt-s}\\
    \nonumber 8\pi G_0 p_r  &=& -\bigl(1+f\bigr) \left(\frac{b}{r^3}-\frac{2\Phi^{'}}{r}\left(1-\frac{b}{r}\right) \right) \\
    &&+ \left(1-\frac{b}{r}\right) \left( \Phi^{'}+\frac{2}{r}\right) f'   \label{IEQ-rr-s} \\
    \nonumber8\pi G_0 p_l  &=& -\bigl(1+f\bigr) \left( \frac{b'r-b}{2r^2} \left(\Phi'+\frac{1}{r} \right)\right. \\
    \nonumber&&-\left. \left(1-\frac{b}{r}\right)\left(\Phi''+\Phi'^2+\frac{\Phi'}{r}\right) \right)\\
     &&+ \left(1-\frac{b}{r}\right)\left( \left( \Phi^{'}+\frac{1}{r}\right)f'+f''\right)-\frac{b'r-b}{2r^2}f' \ , \label{IEQ-pp-s}
  \end{eqnarray}

We will follow exactly the same procedure as in the previous case, considering the Ricci scalar to define our cutoff function,  $f =\xi R$, and consider an equation of state, $p_{r} = \omega(r)\rho$, so that the state parameter $\omega(r)$ is a position-dependent function. Furthermore, we consider a spherical wormhole with zero--tidal force, $\Phi(r) = 0$. With this, we obtain from $p_{r} = \omega(r)\rho$ the following dimensionless equation
\begin{equation}
  \omega \dot{\tilde{b}} u +\tilde{b} = \frac{u^3}{1+f} \left(\left(1-\frac{\tilde{b}}{u}\right)\left(\omega \ddot{f}+\frac{2}{u} (1+\omega)\dot{f}\right)-\omega\frac{\dot{\tilde{b}}u-\tilde{b}}{2u^2}\dot{f}  \right) \, \label{be-s}
\end{equation}
where again we have $u \equiv r/r_{t}$ and $\tilde{b} \equiv b/r_{t}$.

In order to solve the equation \eqref{be-s} we will use the iteration method, as in the previous case. The zeroth order solution is solution of the above equation with the right side being zero, which gives 
\begin{eqnarray}
    \tilde{b}_{(0)}(u)=\omega_t^{\frac{1}{\kappa-\omega_t}} u^{\frac{1}{\kappa-\omega_t}} (\kappa (u-1)+\omega_t)^{\frac{1}{\omega_t-\kappa}}
\end{eqnarray}
with the contour condition $\tilde{b}_{(0)}(1)=1$. Using this $\tilde{b}_{(0)}$, the cutoff function is given by
\begin{equation}
f = \xi R = 2 \zeta  \omega_t^{\frac{1}{\kappa-\omega_t}} u^{\frac{1}{\kappa-\omega_t}-3} [\kappa (u-1)+\omega_t]^{\frac{1}{\omega_t-\kappa}-1},
\end{equation}
which provides for the $\tilde{b}_{(1)}$ equation
\begin{widetext}
\begin{eqnarray}\label{b1sw}
\nonumber\omega_{(0)}\tilde{b}_{(1)}+\omega_{(1)}\dot{\tilde{b}}_{(0)}+\tilde{b}_{(1)} &=& \frac{1}{\omega_t^2 (2 \zeta +\omega_t)^2}\zeta  \left\{\kappa^2 (u-1) \left[10 \zeta +\omega_t (8 \zeta +5 \omega_t+6)\right]+2 \zeta  \kappa \left[2 u (\omega_t+2) (2 \omega_t+1)-\omega_t (5 \omega_t+11)-4\right]+\right.\\
\nonumber&&+\kappa \omega_t \left[2 u (\omega_t (5 \omega_t+9)+3)-\omega_t (11 \omega_t+19)-6\right]+\\
&&\left.+(\omega_t+1) (3 \omega_t+1) \left[(7 u-8) \omega_t^2+2 \zeta  (u (4 \omega_t-1)-5 \omega_t+1)\right]\right\}
\end{eqnarray}
\end{widetext}

Similarly to what was done in the pseudospherical case, in order to properly solve eq. (\ref{b1sw}) for $\tilde{b}_{(1)}$ we must consider the first order expansion of $\omega_{(1)}$ around the throat as $\omega_{(1)}=\omega_T\zeta+\lambda\zeta(u-1)$.

In the spherical case, as well as in the pseudospherical one, the solution for $\tilde{b}_{(1)}$ and consequently for $\tilde{b}(u)$ is expressed in terms of hypergeometric $_2F_1$ functions and are omitted here for the sake of brevity. Thus, the wormhole's traversability will be studied under different configuration of parameters, namely, $(\omega_t, \omega_T, \kappa, \lambda, \zeta)$. As pointed out previously, the wormhole's traversability is studied by analyzing the region of validity of three conditions.

\begin{itemize}
    \item Throat Condition: Since we are considering a zero-tidal model there will be no event horizons and the minimum condition, $\tilde{b}(1)=1$, is satistied by construction. Therefore, only the flare-out condition remains, $\dot{\tilde{b}}(1)<1$, which gives for our model
    
\begin{eqnarray}
-\frac{\zeta  (\kappa+1)+\zeta \omega_t (\kappa+3 \omega_t-\omega_T+6)+\omega_t^2-2 \zeta ^2 \omega_T}{\omega_t^2 (2 \zeta +\omega_t)}<1.
\end{eqnarray}  

\begin{figure}[ht]
    \centering
    \includegraphics[scale=0.9]{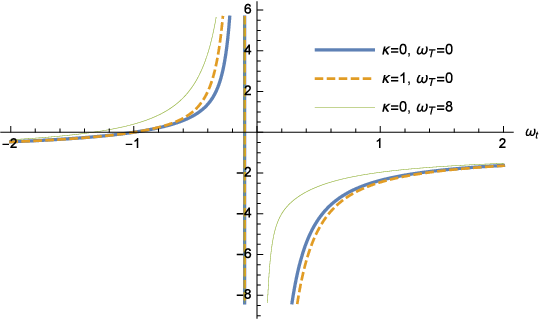}
    \caption{Flare-out condition for spherical wormhole in ASG context with $\zeta=0.05$. Note that in the spherical case the flare-out condition allow the presence of phantom-like matter, and is satisfied for all matter with $\omega_t>1$, including nonexotic matter.}
    \label{fout1_sw}
\end{figure}

Note that, differently from the pseudospherical case, in this case the flare-out condition is satisfied for $\omega_t<-1$, which sets lower limit for phantom-like matter.

\item Antiscreening condition: as already mentioned in the previous section, this condition is related to ASG and requires that $f(\chi)|_{u \rightarrow u_{0}} > 0$, which gives for our model

\begin{eqnarray}
\nonumber\frac{2 \zeta  \left(\zeta \omega_t (\kappa-\omega_T+6)+\zeta (\kappa-2 \zeta  \omega_T+1)+(3 \zeta +1) \omega_t^2\right)}{\omega_t^2 (2 \zeta +\omega_t)}>0\\
\end{eqnarray}

\begin{figure}[ht]
    \centering
    \includegraphics[scale=0.9]{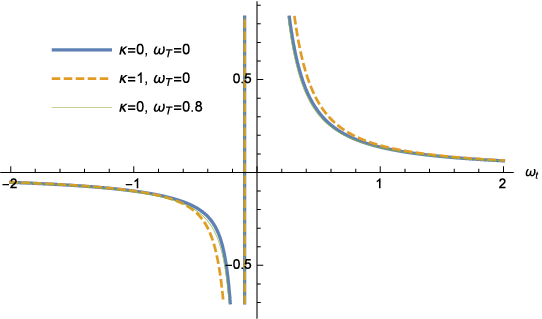}
    \caption{Antiscreening condition for spherical wormhole in ASG context with $\zeta=0.05$. As we can see the antiscreening condition allows the presence of the ordinary matter in the spherical case.}
    \label{asc1_sw}
\end{figure}

\item The null energy condition: The null energy conditions for a zero--tidal model can be written directly from Einstein equations (\ref{IEQ-tt-s}), (\ref{IEQ-rr-s}) and (\ref{IEQ-pp-s}) in which we substitute the shape function $\tilde{b}(u)$ and the cutoff function $f(u)$, yielding $p_r+\rho>0$ as presented in (\ref{neccondsw}). The lateral null energy condition, i.e. $p_l+\rho>0$ was omitted due its length. Unlike the pseudospherical case, in the spherical case the radial and lateral null energy conditions differ significantly one from another, as we can see in figure (\ref{nec1L_sw}). Such difference contributes enormously to the fact that in the spherical case, as we will see later, there are only a few traversable regions.

\begin{figure*}[ht]
\centering
\begin{subfigure}{.3\textwidth}
  \centering
  \includegraphics[scale=0.52]{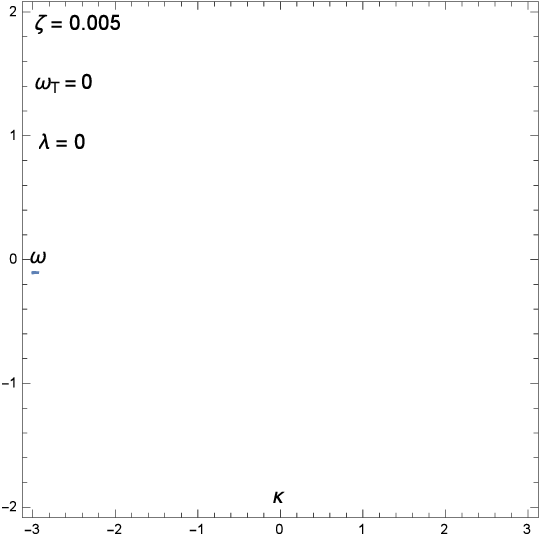}
  \caption{}
\end{subfigure}%
\begin{subfigure}{.3\textwidth}
  \centering
  \includegraphics[scale=0.52]{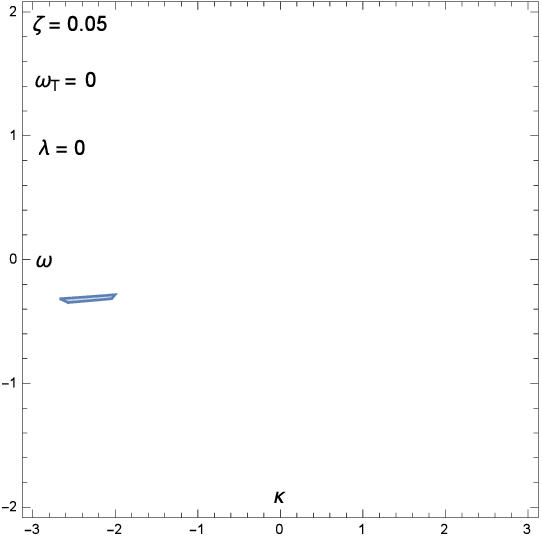}
  \caption{}
 
\end{subfigure}%
\begin{subfigure}{.3\textwidth}
  \centering
  \includegraphics[scale=0.52]{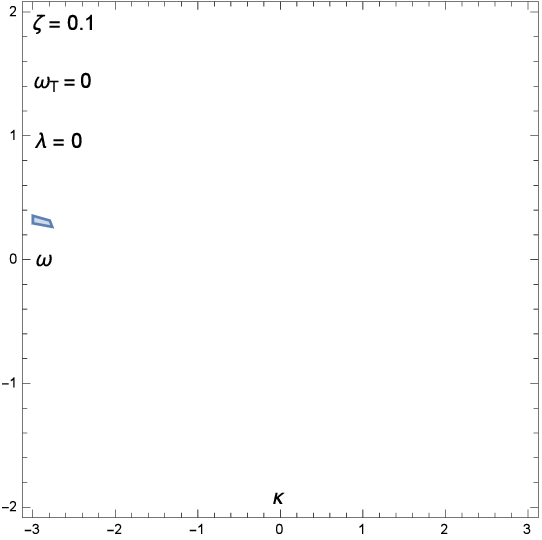}
  \caption{}
 
\end{subfigure}
\caption{In (a) we see the traversability region for spherical worhmholes in the ASG context for $\zeta=0.005$, in (b) we have set $\zeta=0.05$ while in (c) $\zeta=0.1$. In all cases $\omega_T=\lambda=0$. Note that in the spherical case we have the possibility to have traversability with nonexotic matter, only for $\zeta=0.1$ and $\kappa$ around $-3$. This result differs from what was observed in \cite{Moti:2020whf}.}
\label{travsw}
\end{figure*}

Note that for the three conditions in the spherical wormhole case, differently from the pseudospherical case, the divergence appears when $\omega_t=-2\zeta$. However, similarly to the pseudospherical case, the dependence on $\lambda$ is only present on the null energy conditions.

\begin{widetext}
\begin{eqnarray}\label{neccondsw}
\nonumber-\frac{1}{8 \pi  \omega_t^5 (2 \zeta +\omega_t)^3}&&\left[\zeta  \omega_t (\kappa-\omega_T+6)+\zeta  (\kappa-2 \zeta  \omega_T+1)+\omega_t^3+(5 \zeta +1) \omega_t^2\right]\times\\
\nonumber&&\left\{-2 \zeta ^2 \omega_t^2 \left[2 \zeta  (\lambda +11)+3 \kappa^2+\kappa (8 \zeta -\omega_T+14)-(2 \zeta +1) \omega_T+8\right]+\right.\\
\nonumber&&+\zeta ^2 \omega_t \left[-(10 \zeta +7) \kappa^2+4 \kappa (\zeta  (2 \omega_T-9)-2)+2 \zeta  (-2 \zeta  \lambda +2 (\zeta +2) \omega_T-7)-1\right]\\
\nonumber&&\left.-\zeta  \omega_t^3 [14 \zeta  \kappa+\kappa+\zeta  (30 \zeta +\lambda -\omega_T+35)+1]+4 \zeta ^3 (\kappa+1) (2 \zeta  \omega_T-3 \kappa)+\omega_t^5+3 \zeta  (1-8 \zeta ) \omega_t^4\right\}>0
\end{eqnarray}
\end{widetext}

\begin{figure}[ht]
    \centering
    \includegraphics[scale=0.9]{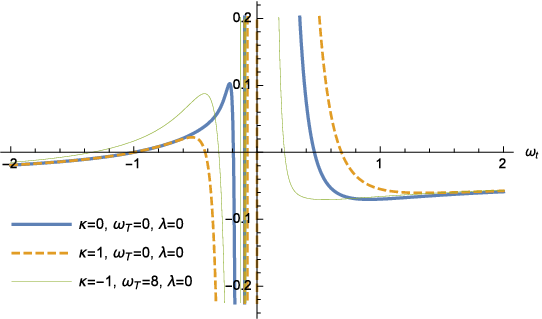}
    \caption{$p_r+\rho$ for spherical wormhole in ASG context with $\zeta=0.05$. In the spherical case the null energy condition have a tiny region with phantom-like matter followed by a quintessence-like matter region. Again, the nonexotic matter is allowed for this condition.}
    \label{nec1_sw}
\end{figure}

\begin{figure}[h!]
    \centering
    \includegraphics[scale=0.9]{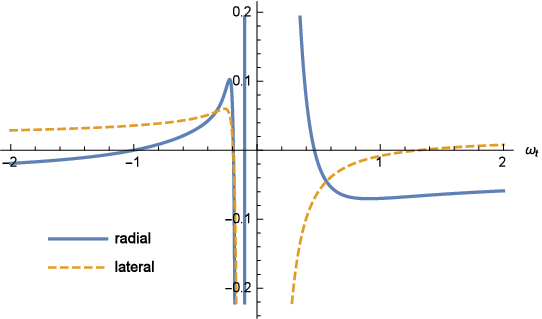}
    \caption{$p_l+\rho$ and $p_r+\rho$ for spherical wormhole in ASG context with $\zeta=0.05$, $\kappa=0$, $\omega_T=0$ and $\lambda=0$. This plot presents a comparison between radial and lateral null energy conditions. We can see that, in the spherical case, both radial and lateral null energy conditions behaves differently.}
    \label{nec1L_sw}
\end{figure}

\end{itemize}

Let us highlight here that under a particular choice of parameters, namely $\omega_t=1$, $\omega_T=8$ and $\kappa=\lambda=0$, we can recover exactly the result found by \cite{Nilton:2021pyi} on the state parameter $\omega$ on the throat in the Ellis-Bronnikov wormhole for small $\xi$ approximation. This result reinforces that the most general result for the zero-tidal wormhole in ASG is achieved by considering $\omega=\omega_{(0)}+\omega_{(1)}$ as considered here.

Finally, analogously to the previous section, in order to study the traversability in the present context, we need to investigate the regions in the space of parameters where all the above conditions are simultaneously satisfied. 

As we can see in figure (\ref{travsw}), we have plotted the region of validity in the space of parameters where all the conditions are simultaneously satisfied at the throat of the wormhole. Analogously to the previous section, we have considered initially $\omega_T=\lambda=0$ in order to disregard the influence of ASG in the state parameter. In fig. (\ref{travsw}a) we have considered $\zeta=0.005$ and we can see that there is a tiny region of traversability with exotic matter $(-1/3<\omega<0)$ when $\kappa\approx-3$. For $\zeta=0.05$ we can see a larger traversable region with exotic matter $(-1/3<\omega<0)$ when $\kappa<0$. And for $\zeta=0.1$ there is also a traversable region with ordinary matter when $\kappa<0$. Similarly to the pseudospherical case, by comparing figures (\ref{travsw}a), (\ref{travsw}b) and (\ref{travsw}c), that the increase in the value of $\zeta$ leads to an increase of traversable regions. 

It is important to point out that, contrary to what is reported in \cite{Moti:2020whf}, in the spherical case it is also possible to have traversability with ordinary matter in the ASG when we consider a non-linear equation of state.

\section{Final remarks}

In this paper we study the traversability of a wormhole in the context of asymptotically safe gravity for two different cases, namely, the spherical case and the pseudospherical case. We carried out an analysis in the throat of the wormhole with an specific choice of the cutoff function $f=\xi R$. We have also considered a non-linear equation of state, so that the state parameter $\omega(u)=\omega_{(0)}(u)+\omega_{(1)}(u,\zeta)$ is written in terms of an ASG independent contribution and an ASG dependent one. 

For both pseudospherical and spherical cases we have obtained the shape function, which are lengthy expressions in terms of hypergeometric $_2F_1$ functions under the consideration of a zero tidal wormhole. In possession of the shape functions we could perform a full analysis of the three conditions that guarantee the wormhole's traversability in the present context, namely, the throat condition, the antiscreening condition and the null energy condition. 

For the pseudospherical wormhole, when $\omega_T=\kappa=\lambda=0$ and $\omega_t\neq0$ the shape function recovers the results presented in \cite{Moti:2020whf}. The flare-out condition excludes the possibility of phantom-like matter at the throat while the other two conditions allow it. We have studied the region of simultaneous validity of the three conditions (fig. (\ref{travpsw})) where we have showed that, exactly on the throat, we only have traversability with non-exotic matter and with exotic ($-1/3<\omega<0$) matter depending on the value of $\kappa$. 

For the spherical wormhole, when $\omega_t=1$, $\omega_T=8$ and $\kappa=\lambda=0$, we recover exactly the result found by \cite{Nilton:2021pyi} on the state parameter $\omega$ on the throat in the Ellis-Bronnikov wormhole for small $\xi$ approximation. Similarly to the analysis performed for the pseudospherical case, all the three conditions were studied separately at first. In this case the flare-out condition allows the presence of phantom-like matter but the antiscreening condition forbids it. The region of simultaneous validity of the three conditions is presented in fig. (\ref{travsw}) where we have a tiny region of traversability with non-exotic matter even when $\kappa\approx-3$, differently from what is reported in \cite{Moti:2020whf}. 

Finally we have showed that the traversable regions exhibit, in both cases, a strong dependence with the $\zeta$ parameter. Such dependence is expected, since the greater is the $\zeta$ the greater is the quantum influence in the wormhole. A natural continuation of this work consists in the consideration of other choices for the cutoff function, such as the squared Ricci and the Kretschmann scalar.

\section*{Acknowledgements}

G.A. would like to thank Conselho Nacional de Desenvolvimento Científico e Tecnológico (CNPq) and Fundação Cearense de Apoio ao Desenvolvimento Científico e Tecnológico
(FUNCAP) and M.N. would like to thank Coordenação de Aperfeiçoamento de Pessoal de Nível Superior - Brasil (CAPES) for finantial support

\appendix

\section{shape functions}

The shape function for the pseudospherical wormhole case is written as
\begin{widetext}
\begin{eqnarray}
\nonumber \tilde{b}(u)&=&\frac{1}{\omega _t^2 \left(\kappa -\omega _t\right)^2}\left\{\frac{\zeta  \left(\kappa -\omega _t\right)}{\left(\omega _t-2 \zeta \right){}^2}\left[\omega _t^{\frac{1}{\kappa -\omega _t}+1} \, _2F_1\left(1,1,1+\frac{1}{\omega _t-\kappa },\frac{\kappa }{\kappa -\omega _t}\right) u^{\frac{1}{\kappa -\omega _t}} \left(\omega _t+\kappa  (u-1)\right){}^{\frac{1}{\omega _t-\kappa }}\right.\right.\times\\
\nonumber&&\times\left[\kappa ^2 \left(10 \zeta +\omega _t \left(8 \zeta -5 \omega _t-6\right)\right)+\kappa  \left(8 \zeta +\omega _t \left(22 \zeta +\omega _t \left(10 \zeta -11 \omega _t-19\right)-6\right)\right)-2 \left(\omega _t+1\right) \left(3 \omega _t+1\right) \left(\zeta -5 \zeta  \omega _t+4 \omega _t^2\right)\right]+\\
\nonumber&&+\frac{1}{\kappa}\left[\left(\kappa -\omega _t\right) \left(\omega _t+\kappa  (u-1)\right) \, _2F_1\left(1,1,1+\frac{1}{\omega _t-\kappa },\frac{u \kappa }{\kappa -\omega _t}\right)\left(\kappa ^2 \left(\omega _t \left(5 \omega _t+6\right)-2 \zeta  \left(4 \omega _t+5\right)\right)\right.\right.+\\
\nonumber&&+\left.\left(\omega _t+1\right) \left(3 \omega _t+1\right) \left(2 \zeta -8 \zeta  \omega _t+7 \omega _t^2\right)+2 \kappa  \left(\omega _t \left(\omega _t \left(5 \omega _t+9\right)+3\right)-2 \zeta  \left(\omega _t+2\right) \left(2 \omega _t+1\right)\right)\right)+\\
\nonumber&&+\frac{\omega _t}{-\kappa +\omega _t+1}\left[\kappa  \omega _t^{\frac{1}{\kappa -\omega _t}} \, _2F_1\left(1,1,2+\frac{1}{\omega _t-\kappa },\frac{\kappa }{\kappa -\omega _t}\right) u^{\frac{1}{\kappa -\omega _t}} \left(\omega _t+\kappa  (u-1)\right){}^{\frac{1}{\omega _t-\kappa }}+\left(\kappa -\omega _t-1\right) \left(\kappa -\omega _t\right)\right]\times\\
\nonumber&&\times\left(\kappa ^2 \left(\omega _t \left(6 \omega _t+7\right)-2 \zeta  \left(5 \omega _t+6\right)\right)+\kappa  \left(\omega _t \left(\omega _t \left(13 \omega _t+22\right)+7\right)-2 \zeta  \left(7 \omega _t \left(\omega _t+2\right)+5\right)\right)\right.+\\
\nonumber&&+\left.\left.\left.\left(\omega _t+1\right) \left(3 \omega _t+1\right) \left(2 \zeta -8 \zeta  \omega _t+7 \omega _t^2\right)\right)\right]\right]+\frac{\omega_t}{\omega _t+\kappa  (u-1)}\times\\
\nonumber&&\times\left[2 u \omega _t \left(\kappa -\omega _t\right)^2 \left(\omega _t+\kappa  (u-1)\right)\omega _t^{\frac{1}{\kappa -\omega _t}} \left(-u^{\frac{1}{\kappa -\omega _t}}\right) \left(\omega _t+\kappa  (u-1)\right){}^{\frac{1}{\omega _t-\kappa }}\right.\times\\
\nonumber&&\times\left(\zeta  \left(-\log \left(\omega _t+\kappa  (u-1)\right)+\log \left(\omega _t\right)+\log (u)\right)\omega _t \left(\omega _T-\lambda \right) \left(\omega _t+\kappa  (u-1)\right)\right.\\
&&+\left.\left.\left.\left(\kappa -\omega _t\right) \left(-\omega _t \left(\omega _t^2+\zeta  \lambda  (u-1)\right)+\kappa ^2 (u-1) \omega _t-\kappa  (u-2) \omega _t^2+\zeta  \kappa  (u-1) \omega _T\right)\right)\right]\right\}
\end{eqnarray}
\end{widetext}

The shape function for the spherical wormhole case is

\begin{widetext}
\begin{eqnarray}
\nonumber\tilde{b}(u)&=&\frac{1}{\omega _t^2 \left(\kappa -\omega _t\right)^2}\left\{\frac{\zeta  \left(\kappa -\omega _t\right)}{\left(2 \zeta +\omega _t\right)^2}\right.\left[-\omega _t^{\frac{1}{\kappa -\omega _t}+1} \, _2F_1\left(1,1,1+\frac{1}{\omega _t-\kappa },\frac{\kappa }{\kappa -\omega _t}\right) u^{\frac{1}{\kappa -\omega _t}}\left(\omega _t+\kappa  (u-1)\right)^{\frac{1}{\omega _t-\kappa }}\right.\times\\
\nonumber&&\times\left[\kappa ^2 \left(10 \zeta +\omega _t \left(8 \zeta +5 \omega _t+6\right)\right)+\kappa  \left(8 \zeta +\omega _t \left(22 \zeta +\omega _t \left(10 \zeta +11 \omega _t+19\right)+6\right)\right)+2 \left(\omega _t+1\right) \left(3 \omega _t+1\right) \left(-\zeta +5 \zeta  \omega _t+4 \omega _t^2\right)\right]\\
\nonumber&&+\frac{1}{\kappa}\left[\left(\kappa -\omega _t\right) \left(\omega _t+\kappa  (u-1)\right) \, _2F_1\left(1,1,1+\frac{1}{\omega _t-\kappa },\frac{u \kappa }{\kappa -\omega _t}\right)\left(\kappa ^2 \left(10 \zeta +\omega _t \left(8 \zeta +5 \omega _t+6\right)\right)+\right.\right.\\
\nonumber&&+\left.\left(\omega _t+1\right) \left(3 \omega _t+1\right) \left(-2 \zeta +8 \zeta  \omega _t+7 \omega _t^2\right)+2 \kappa  \left(4 \zeta +\omega _t \left(10 \zeta +\omega _t \left(4 \zeta +5 \omega _t+9\right)+3\right)\right)\right)+\\
\nonumber&&+\frac{\omega _t}{-\kappa +\omega _t+1}\left[\kappa  \omega _t^{\frac{1}{\kappa -\omega _t}} \, _2F_1\left(1,1,2+\frac{1}{\omega _t-\kappa },\frac{\kappa }{\kappa -\omega _t}\right) u^{\frac{1}{\kappa -\omega _t}} \left(\omega _t+\kappa  (u-1)\right)^{\frac{1}{\omega _t-\kappa }}+\left(\kappa -\omega _t-1\right) \left(\kappa -\omega _t\right)\right]\times\\
\nonumber&&\times\left.\left(\kappa ^2 \left(2 \zeta  \left(5 \omega _t+6\right)+\omega _t \left(6 \omega _t+7\right)\right)+\kappa  \left(2 \zeta  \left(7 \omega _t \left(\omega _t+2\right)+5\right)+\omega _t \left(\omega _t \left(13 \omega _t+22\right)+7\right)\right)\right.\right.\\
\nonumber&&\left.\left.\left.+\left(\omega _t+1\right) \left(3 \omega _t+1\right) \left(-2 \zeta +8 \zeta  \omega _t+7 \omega _t^2\right)\right)\right]\right]+\omega _t^{\frac{1}{\kappa -\omega _t}+1} u^{\frac{1}{\kappa -\omega _t}} \left(\omega _t+\kappa  (u-1)\right)^{\frac{1}{\omega _t-\kappa }-1}\times\\
\nonumber&&\times\left[\zeta  \omega _t \left(\omega _T-\lambda \right) \left(\omega _t+\kappa  (u-1)\right) \left(-\log \left(\omega _t+\kappa  (u-1)\right)+\log \left(\omega _t\right)+\log (u)\right)\right.\\
&&\left.\left.+\left(\kappa -\omega _t\right) \left(-\omega _t \left(\omega _t^2+\zeta  \lambda  (u-1)\right)+\kappa ^2 (u-1) \omega _t-\kappa  (u-2) \omega _t^2+\zeta  \kappa  (u-1) \omega _T\right)\right]\right\}
\end{eqnarray}
\end{widetext}

\end{document}